\documentstyle[12pt]{article}
\def\be{\begin{equation}}
\def\ee{\end{equation}}
\def\bea{\begin{eqnarray}}
\def\eea{\end{eqnarray}}
\def\s{\sigma}
\def\al{\alpha}

\def\de{\delta}
\def\om{\omega}
\title{Hadronic Regge Trajectories:\\
 Problems and Approaches}

\author{A. Inopin \\
\em\small Virtual Teacher Ltd., Vancouver, Canada; E-mail:
inopin@yahoo.com \\
 \\
G.\,S. Sharov \\
\em\small Tver state university, \\
\em\small 170002, Sadovyj per., 35, Tver, Russia; E-mail: german.sharov@tversu.ru}
\date{}
\begin{document}
\maketitle
\begin{abstract}

We scrutinized hadronic Regge trajectories in a framework of two different
models --- string and potential. Our results are compared with broad spectrum
of existing theoretical quark models and all experimental data from PDG98.
It was recognized that Regge trajectories for mesons and baryons are not
straight and parallel lines in general in the current resonance region
both experimentally and theoretically, but very often have appreciable curvature, which is flavor-dependent. For a set of baryon Regge trajectories
this fact is well described in the considered potential model.
The standard string models predict linear trajectories at high angular
momenta $J$ with some form of  nonlinearity at low $J$.

\end{abstract}

\medskip
\noindent{\large\bf Introduction}
\medskip

In the past few years Regge trajectories (RT) have attracted the interest of many authors,
who build quark models of baryons, mesons, glueballs and hybrids [1\,--\,19].
Quite a number of approaches have been used to attack this problem:
the WKB-approach \cite{Martin,FR}, the Wilson loop model \cite{SimFR,Sim}, the $\hbar$-expansion technique \cite{Step}, q-deformed algebra approach \cite{Dey},
$1/N_c$-approach \cite{t'H}, spectrum generating algebra (SGA) model \cite{Iach},
Filipponi model \cite{Filip}, NRQM \cite{Inop}, the string models
\cite{Ch}\,--\,\cite{4B}, extended covariant oscillator quark model for glueballs
\cite{stringlue},
$N/D$-method \cite{N/D} and others.

Probably, the most interesting questions, which are under investigation, are the following: (a) Are the RT really straight lines in the entire energy interval, or this is only true asymptotically? (b) Do the trajectories for mesons, baryons, glueballs and hybrids have the same slope? (c) What is the flavor-dependence of RT? (d) What is the intrinsic connection between kinematics, the type of the potential and the straight  RT? (e) Is it possible to connect the linear RT with parity-doubling phenomena (tower structure)?
(f) What is the effect of color screening on heavy and light quarkonia and baryon's RT? (g) When does the asymptotic regime (J=?) really start for baryons, mesons, glueballs, hybrids for both parent and daughter (ancestor) trajectories? (h) What is the dependence of the character of RT on the scalar/vector structure of the confinement potential?

Different groups pursued variety of models and different aspects of this
problem. It has been established long time ago, that the experimental RT for N,
$\Delta$ baryons are not strictly straight lines \cite{HeyK}.
Authors of this seminal review and Hendry in review \cite{Hendry} considered the
facts of nonlinear behavior of the RT in mass-squared. Rather, as Hendry concludes, baryon resonances seem more linear as a function of cm momentum.

In this paper we will show that both experimentally and theoretically RT may
deviate
from straight and parallel lines, depending on particular family of baryons, mesons, glueballs or hybrids and energy region. We will scrutinize this problem in a framework of two different models --- string \cite{4B},
and potential \cite{Inop}.
Then we will contrast our results with broad spectrum of existing theoretical models and experimental data for different flavors and draw the conclusions.

\bigskip
\noindent{\large\bf 1. Regge Trajectories in the Potential Quark Model}
\medskip

In our recent series of papers \cite{Inop}, based on hamiltonian (1) and the method of hyperspherical functions (HF) \cite{RichT}, a description of $N$, $\Delta$, $\Omega$ resonance spectra and partial widths was given.
\bea
&H=H_0+H_{hyp},&\label{H}\\
&\displaystyle H_0=\sum_{i=1}^3m_i+\sum_{i=1}^3\frac{\mbox{\boldmath$ P$}_i^2}
{2 m_i }-\frac23\sum_{i<j}\left(\frac{\al_s}{r_{ij}}-br_{ij}\right)+V_0,&\nonumber\\
&\displaystyle H_{hyp}=\frac23\sum_{i<j}\frac{\pi C_\al}{m_i^2}\left(1+\frac83
(\mbox{\boldmath$s$}_i\mbox{\boldmath$s$}_j) \right)\,
\de(\mbox{\boldmath$r$}_{ij})+\sum_{i<j}\frac{2C_t}{3m_i m_j r_{ij}^3}\left[
3\frac{(\mbox{\boldmath$s$}_i\mbox{\boldmath$r$}_{ij})(\mbox{\boldmath$s$}_j
\mbox{\boldmath$r$}_{ij})}{r_{ij}^2}-\mbox{\boldmath$s$}_i\mbox{\boldmath$s$}_j
\right].&\nonumber
\eea

Following Ref.~\cite{CutkF}, we introduce the constants $\al_s$,
$C_\al$ and $C_t$, which determine the strength
of the Coulomb, contact and tensor potentials respectively.
The use of the hamiltonian (1) allows us to obtain
better agreement with experiment for resonances of positive
and negative parity, and also to describe resonances
with both large J and M, and with small J and M.

We showed \cite{Inop} that it was  appropriate for such
a description to take advantage of the concept of yrast states
and yrast lines from the theory of atomic nuclei rotational spectra,
as well as to make use of the concept of RT.
One of the main results was that both theoretical and experimental
spectra are {\it nonlinear} trajectories in Chew-Frautschi plots
throughout the experimental region. Our model predicts a whole
series of high-lying $N$, $\Delta$ resonances, represented in Table 1.

\begin{center}
Table 1. Masses of the predicted N, $\Delta$ resonances in the model \cite{Inop}\\
\smallskip

\begin{tabular}{|c|c|c|} \hline
\makebox[10em]{N $11/2^+$ (2.39)}& \makebox[10em]{N $13/2^-$ (2.90)}&
\makebox[10em]{N $15/2^+$ (3.00)}\\ \hline
N $17/2^-$ (3.45)& N $19/2^+$ (3.45)& N $21/2^-$ (3.95)\\ \hline
$\Delta$ $15/2^-$ (3.48)& $\Delta$ $17/2^+$ (3.49)& $\Delta$ $19/2^-$ (4.00)\\
\hline
\end{tabular}
\end{center}

The present paper generalize the model \cite{Inop} to u, d,
s, c, b flavors (it is impossible to create top baryons
and mesons) and a wider range of angular momenta $L=0\div20$.
We obtain the spectra of BR in the same
way as in \cite{Inop},  by solving Schr\"odinger equation
(SE) with the HF method.

When the hadron wave function (WF) is expanded in the HF
basis and substituted in the SE, one generally finds
an infinite system of differential equations for the radial
WF (RWF). However, as was shown in \cite{RichT}
for a system of identical u,d-quarks, the coupling of
channels is weak, therefore it is sufficient to include only
several terms in $K$, grand orbital momentum, in the WF expansion.
With increasing excitation energy,
the contribution of $H_{hyp}$ vanishes, hence the coupling
of channels in general can be neglected.
On the other hand, as we will consider only symmetric
 baryons with increasing quark mass (u, s, c, b),
the above mentioned arguments will apply even more strongly.

Now, let us introduce Jacobi and hyperspherical coordinates
 for a three-body problem. The Jacobi coordinates
are defined as usual (please see \cite{Inop} for all the details)
\be
\mbox{\boldmath$R$}=(\mbox{\boldmath$r$}_1+\mbox{\boldmath$r$}_2+
\mbox{\boldmath$r$}_3)/3,\quad
\mbox{\boldmath$\eta$}=(\mbox{\boldmath$r$}_1-\mbox{\boldmath$r$}_2)/
\sqrt2,\quad\mbox{\boldmath$\xi$}=(\sqrt{2/3})\big[(\mbox{\boldmath$r$}_1+
\mbox{\boldmath$r$}_2)/2-\mbox{\boldmath$r$}_3 \big].
\label{R}\ee

The hyperspherical angle $\theta$ and radius $\rho$ are defined as follows
\be
\eta=\rho\cos\theta,\quad\xi=\rho\sin\theta,\quad 0<\theta<\pi/2,\quad
\rho=(\eta^2+\xi^2)^{1/2},\quad 0<\rho<\infty.
\label{eta}\ee
So, our WF and q-q potentials will depend on new variables, {\boldmath$\eta$
and $\xi$}.

We will work in the {\it hypercentral} approximation (HCA), where only that part of the interaction which is
invariant under rotation in six-dimensional space $(\rho,\Omega_5)$ is taken into account. Because in HCA
baryons are pure {\it rotational} states in six-dimensional space, and because a very soft potential is used in our
model, no strong correlations generating a so-called quark-diquark state can occur. So, we will neglect
the $\theta$-dependence of our q-q potential. The introduction of $\theta$-dependence in the potential mixes
 the states belonging to one value of $L$, and slightly shifts their positions in the baryonic mass spectrum. As
was proved by Richard \cite{RichT}, the most commonly used potentials in hadron spectroscopy (including ours)
are very close to hypercentral (please see \cite{Inop} for all the details).

\bigskip
\noindent{\large\bf 2. Mass Spectra and RT in the Potential Quark Model}
\medskip

Now let us discuss our results for mass spectra and RT of baryons. The method of numerical  solution of the SE
was described in detail in \cite{Inop}, but we will briefly recapitulate it here. The system of differential
equations (SDE) was reduced to a system of first -order differential equations, which we integrated by
determining a full set of Cauchy solutions. We then constructed the required solution by imposing the
following boundary conditions on the RWF: $F_i(\rho=0)=0$ and $F_i(\rho=R_\infty)=0$. The eigenvalues (EV)
were determined by zeroing the determinant constructed from the fundamental system of solutions for SDE.
For the parameter $R_\infty$  we always choose a much larger value than the radius of the corresponding excited
state $\langle\rho\rangle_i$, which grows with $L$ and $N_r$ (the radial quantum number). We always choose
$R_\infty$ such that if we take $R_\infty=2R_\infty$, then the corresponding EV of any excited state will not change
more than 1\%. We use in our computation the following set of parameters (Table 2). With this input we have
calculated the mass spectra, RT, mean radii $\langle\rho\rangle_i$ and slopes for u, d, s, c, b flavors,
$N_r=0,\,1,\,2$ and momenta range $L=0\div20$.

\begin{center}
Table 2. Input parameters in the model \cite{Inop}\\
\smallskip

\begin{tabular}{|l|l|l|} \hline
$\al_s=C_\al=1$ &     $C_t=0.2$& $ b=350$ MeV/fm\\ \hline
$V_0=-513$ MeV&   $m_u=330$ MeV&  $m_d=330$ MeV\\ \hline
$m_s=607$ MeV &   $m_c=1500$ MeV&  $m_b=5170$ MeV\\ \hline
\end{tabular}
\end{center}

We note that the solution of the SE in the single-channel approximation with
the centrifugal potential $V_c=(K+3/2)(K+5/2)\big/(2m\rho^2)$ is equivalent
from the mathematical point of view to the Davydov-Chaban
model \cite{DavyC}, in which the rotational spectrum of a {\it non-spherical nucleus} with variable moment of inertia
is calculated. Because the moment of inertia is not constant but varies {\it linearly} with the angular momentum $J$
of the nucleus, the rotational spectrum, instead of the {\it quadratic} behavior $J(J+1)$, it would have for a constant
moment of inertia, approaches one {\it linear} in $J$.

Approximately the same thing happens with the hadron: for large excitation energies the spectrum becomes {\it linear}
in $J$, and the hadron becomes a strongly extended system along its {\it symmetry axis}. This should lead to a linear
nature of the baryonic mass spectra along yrast lines. Our predictions are in accord with findings by Hey \cite{HeyK}
and Hendry \cite{Hendry}.

Let us start a detailed, sector by sector comparison of our results for
different flavors.  Note that RT ($M^2=M^2(J)$) for baryons have different
types of curvatures: some are convex; others are concave functions of $J$
(see Figs.~1,\,2). The u-d (N, $\Delta$) trajectories are {\it concave} for
$N_r=0,\,1,\,2$ (Fig.~2), whereas strange (sss) RT for $N_r=0,\,1,\,2$ are
"stereotypical" --- they are nearly straight lines with slowly varying slopes
$\al'$ (Fig.~3). Charmed (ccc) and bottom (bbb) trajectories are
{\it convex} for $N_r=0,1,2$.

When we analyze the mass spectra $ M=M(J)$ for u-d, s, c, b-baryons, they are all
convex functions of $J$, with a rather complex dependence on
$m_q$ (flavor) and $N_r$.

If we fix $m_q$ and look at the Chew-Frautschi plot for $N_r=0,\,1,\,2$, we
can see that these three curves generally are {\it non-parallel} (Fig.~2),
that strongly contradicts the conventional picture of the parallel RT. It is
noteworthy that Olsson et. al. \cite{Olsson} recently examined the meson
sector, using a relativistic flux-tube model, and noticed nonlinearity of RT
at low angular momenta $J$.  But it remains unclear whether their model
accounts adequately for physical observables, because it describes a very
limited number of states in the mesonic sector.

Slopes for the u-d family start above 1 GeV${}^{-2}$, with the largest value
for $N_r=2$; the curves decrease almost monotonically with changing
the rate of decreasing near $L=4$, for $N_r=0,1,2$ (see Fig.~3).

Slopes for the s-family differ from all the other cases, because they are
rather {\it weak} functions of $L$. Slopes for daughters $N_r=1,\,2$ start
just above unity, then slowly decrease to 0.87 GeV${}^{-2}$ and 0.78
GeV${}^{-2}$, respectively, whereas the parent slope {\it fluctuates} slightly
over the interval, spanning the range 0.97-0.91 GeV${}^{-2}$ (see Fig.~3).

Slopes for the charm family are highly {\it nonlinear} functions of $L$. The
$N_r=0,\,1,$ slopes increase monotonically, starting from 0.39 GeV${}^{-2}$
and 0.55 GeV${}^{-2}$ and approaching 0.889 GeV${}^{-2}$ and 0.893
GeV${}^{-2}$, respectively. The $N_r=2$ slope has a dip at $L=4$, and grows
continuously, reaching 0.897 GeV${}^{-2}$.  These values are almost {\it
identical} to Olsson's results for mesons for asymptotic $L$ \cite{Olsson}.

Bottom baryon slopes differ {\it sharply} from the u-d-s-c
sector, first, by their small magnitude, and second, by the significant
increase of the slopes  along the trajectory (the B-slopes increase by about an
order of magnitude, from 0.055 GeV${}^{-2}$ to
0.504 GeV${}^{-2}$). Daughter slopes for $N_r=2$ fluctuate, but still
increase with $L$.

For convenience, we present the set of median values $\langle\al'\rangle_i$  for the whole flavor multiplet
(see, Table 3).  It is interesting to note that the median values of $\al'_i$ are almost independent on $N_r$ for the
u-d and s-families.

\begin{center}
Table 3. Median values $\langle\al'\rangle_i$ for trajectories with $N_r = 0,\,1,\,2$            \\
\smallskip

\begin{tabular}{|c|c|c|c|c|} \hline
$N_r$& Up-Down&   Strange&   Charm&   Bottom\\ \hline
  0&     0.90&     0.95&     0.73&     0.31\\ \hline
  1&     0.87&     0.95&     0.77&     0.34\\ \hline
\makebox[5em]{2}&\makebox[7em]{0.84}&\makebox[7em]{0.94}&\makebox[7em]{0.81}&
\makebox[7em]{0.37}\\
\hline
\end{tabular}
\end{center}

The expectation values of the hyperradius $\langle\rho\rangle_i$ are basically smooth increasing functions of $L$
and $N_r$, and decreasing functions of $m_q$.

We proved that the slope of the trajectories decreases with increasing quark mass in the mass region of the {\it lowest}
excitations.  This is due to the contribution of the color Coulomb interaction, which increases with mass, leading to a
{\it curvature} of the trajectory near the ground state.  In the asymptotic regime, the trajectories for all flavors are linear
and have the same slope $\al'\simeq$ 0.9 GeV${}^{-2}$.

After careful numerical evaluation of baryonic and mesonic spectra for all flavors we have shown that the point
of establishment of linear RT depends on the exponent $\nu$ in the power law potential in hamiltonian (1), and occurs
at larger $L$ with larger $\nu$.  For an oscillator confinement ($\nu=2$) linear regime started only from $L>20$,
for linear confinement ($\nu=1$) linear regime started from $L>18$.

We proved that for mesons linear RT is established earlier than for baryons as a function of $L$.  The reason lies in
{\it centrifugal energy} term, which has $L(L+1)$ dependence for mesons and $(L+3/2)(L+5/2)$ dependence
for baryons.  As we see, {\it three-dimensional} corrections to $L(L+1)$-law are important for baryons.

\bigskip
\noindent{\large\bf 3. String models}
\medskip

The string models are widely used for describing orbitally excited
hadron states by virtue of some remarkable features: (a) the direct analogy
between the string with linearly growing energy and the QCD confinement
mechanism of connecting quarks (antiquarks) by the gluon field tube \cite{BN};
(b) the strings are relativistic by definition\footnote{The relativistic string
dynamics results from the extremization of a world surface area swept by the
string in Minkowski space.}; (c) the energy $E=M$ and the angular momentum
$J$ of a rotating open (massless) string are connected by the Nambu
relation \cite{Nambu}
\be
J=\al'M^2,\qquad \al'=(2\pi\gamma)^{-1},
\label{jnam}\ee
where $\gamma$ is the string tension. This fact allows us to apply
the string models to describing the RT.

The massless string generates the strictly linear RT (\ref{jnam}).
But the meson model of relativistic string with massive ends \cite{Ch}
(more realistic one in comparison with the massless case) results in the
more complicated behavior of RT. For this model the relation (\ref{jnam})
takes place only in the high energy limit \cite{Ch,Ko}.

Shortly after this meson model string models of baryon were suggested
in four variants differing from each other in the topology of spatial
junction of three massive points (quarks) by relativistic strings:
(a) the quark-diquark model q-qq \cite{Ko} (from the point of view
of classical dynamics it coincides with the mentioned meson model of
relativistic string with massive ends \cite{Ch,BN});
(b) the linear configuration q-q-q with quarks connected in series
\cite{4B,lin};
(c) the ``three-string" model or Y-configuration with three
strings joined in the fourth massless point (junction) \cite{AY,PY}
and (d) the ``triangle" model or $\Delta$-configuration that could be
regarded as a closed string carrying three pointlike masses \cite{Tr,PRTr}.

All mentioned string hadron models may be constructed
on the base of the action \cite{Ch,4B,BN}
\be
S=-\int\limits_{\tau_1}^{\tau_2}\!\biggl\{\gamma\!
\int\limits_{\s_*(\tau)}^{\s_N(\tau)}\!\!
\sqrt{(\dot XX')^2-\dot X^2X'{}^2}\,d\s+
\sum_{i=1}^Nm_i\sqrt{V_i^2(\tau)}\biggr\}d\tau.
\label{S}\ee
Here $m_i$ are masses and $N$ is the number of the material points
(quarks, antiquarks or diquarks)\footnote{For brevity we will use the term
``quark" instead of ``material point", keeping in mind that
the action (\ref{S}) on the classic level doesn't describe the spin
and other quantum numbers of quarks.}, $N=2$ for the meson and quark-diquark
models and $N=3$ for other models of baryon,
$x^\mu=X^\mu(\tau,\s)$ is the string world surface
in $d$-dimensional Minkowski space with signature $+,-,-,\dots$,
$\,\dot X^\mu=\partial_\tau X^\mu$,
$X'{}^\mu=\partial_\s X^\mu$, $c=1$, $\hbar=1$;
$V_i^\mu=\frac d{d\tau}X^\mu(\tau,\s_i(\tau))$ is the tangent
vector to the i-th quark trajectory $\s=\s_i(\tau)$,
$\s_i(\tau)<\s_{i+1}(\tau)$, $\s_*\equiv\s_1(\tau)$
for all models, except for the $\Delta$ configuration.

In the ``triangle" model the equations $\s=\s_*=\s_0(\tau)$
and $\s=\s_3(\tau)$ define the trajectory of the same (the 3-rd)
quark. This fact may be written as the closure condition
$X^\mu(\tau,\s_0(\tau))=X^\mu(\tau^*,\s_3(\tau^*))$ \cite{Tr}.
The parameters $\tau$ and $\tau^*$ aren't equal in general.

The equations of motion and the boundary conditions on the quark
trajectories in these models are deduced by variation and minimization
of action (\ref{S}).
In particular, under the conditions of orthonormality $\dot X^2+X'{}^2=0$,
$(\dot XX')=0$ (they may be obtained for all configurations \cite{BN,Tr})
the equations of motion become linear
\be
\ddot X^\mu-X''{}^\mu=0,
\label{eq}\ee
and the boundary conditions take the simplest form
\be
m_i\frac d{d\tau}\frac{V_i^\mu}{\sqrt{V_i^2}}=F_i^\mu.
\label{qq}\ee
Here $F_i^\mu=\pm\gamma\big[X'{}^\mu+\s_i'(\tau)\,\dot X^\mu\big]
\Big|_{\s=\s_i+0}$ for a quark at an end of the string,
this expression takes another form
$F_i^\mu=\gamma\big[X'{}^\mu+\s_i'(\tau)\,\dot X^\mu\big]
\Big|_{\s=\s_i+0}\!\!=\gamma\big[X'{}^\mu+\s_i'(\tau)\,\dot X^\mu\big]
\Big|_{\s=\s_i-0}$
for a quark in the ``triangle" model and for the middle
quark in the q-q-q configuration \cite{Tr}. On this trajectories
the derivatives of $X^\mu(\tau,\s)$ are not continuous in general.

The string models are successfully applied to describing the main
or parent RT. For this RT, that is for the orbitally excited hadrons
the rotational motions of all string configurations (flat uniform rotations
of the system) are used \cite{Ko,4B,PRTr,Solov}.

The solution of this type satisfying Eq.~(\ref{eq}), conditions
(\ref{qq}) and describing the uniform rotation of the rectilinear string
is well known for the meson string model \cite{Ch,BN} (or its equivalent
q-qq model) and for the q-q-q baryon configuration.
It may be represented as ($N=3$ for the q-q-q)
\be
X^0\equiv t=\tau,\qquad
X^1+iX^2=\om^{-1}\sin\om\s\cdot e^{i\om\tau}.
\label{sol}\ee
Here $\om$ is the angular velocity, $X^1\equiv x$, $X^2\equiv y$,
$\s\in[\s_1,\s_N]$, $\s_i={}$const, $\s_1<0$, $\s_N>0$.

For the linear q-q-q configuration the middle quark is at rest at
$\s=\s_2=0$. But this motion is unstable with respect to centrifugal
moving away of the middle quark that results in the complicated
quasi-periodical motion with varying distance between the nearest two quarks
\cite{lin}. However the minimal value of the mentioned distance
does not equal zero, in other words, the system q-q-q is not transformed
in the  quark-diquark one as was supposed in Refs.~\cite{Ko}.
So the q-q-q system is probably applicable not to pure orbital excitations
but to radial ones.

The rotational motion of the ``three-string" model with the junction at rest
and with rectilinear string segments joined in a plane of rotation at the
angles 120${}^\circ$ \cite{AY,PY} is described by the expression similar to
Eq.~(\ref{sol}).

For the meson string model with massive ends and for the baryonic
configurations q-qq, q-q-q and Y the energy $E\equiv M$ and the angular momentum
$J$ ($z$ projection) of the considered rotational motions are \cite{Ch,Ko,4B}
\bea
M&=&\sum_{i=1}^N\bigg[\frac\gamma\om\arcsin v_i+
\frac{m_i}{\sqrt{1-v_i^2}}\bigg]+\Delta M,\label{M}\\
J&=&\sum_{i=1}^N\left[\frac1{2\om}\bigg(\frac\gamma\om
\arcsin v_i+\frac{m_iv_i^2}{\sqrt{1-v_i^2}}\bigg)+s_i\right],
\label{J}\eea
where the velocities of moving quarks
$v_i=\sin|\om\s_i|=\sqrt{\big(\frac{m_i\om}{2\gamma}\big)^2+1}-
\frac{m_i\om}{2\gamma}$.
The presence of quark spins with projections $s_i$ ($\sum_{i=1}^Ns_i=S$)
is taken into account as the correction $\Delta M$ to the energy of
the classic motion.
In Refs.~\cite{Ko,4B} this correction is due to the spin-orbit interaction
(the spin-spin correction is assumed to be small in comparison with the
spin-orbit one at high $J$).

L.\,D. Soloviev \cite{Solov} in the frameworks of another approach constrains
the string dynamics only rectilinear string motions (\ref{sol}) (with motions of the c.m.)
and introduces into action (\ref{S}) the quark spin terms with some additional terms.
The latter ones compensate the influence of the quark spins so that
the resulting motion is the same as in the spinless case. Such a form
of compensation results in Ref.~\cite{Solov} in the same expressions (\ref{M})
and (\ref{J}) with $\Delta M=0$ and two corrections in Eq.~(\ref{J}).
The l.~h.~s. of Eq.~(\ref{J}) in Ref.~\cite{Solov} is substituted by
$\sqrt{J(J+1)}$ due to some form of quantization and the spin term
$S=\sum_{i=1}^Ns_i$ is substituted by the phenomenological parameter $a_n$
resulting from the quark interaction at short distances. The values $a_n$
(different for various RT and depending on $J$ for heavy quarks) are
calculated by fitting.

Using this approach L.\,D. Soloviev \cite{Solov} describes a lot of meson
states, in particular, the states with zero orbital momentum ($\pi$, $\rho$),
where the string models are unlikely adequate.

The rotational motions (uniform rotations about the system center of mass) for
the ``triangle" baryon model \cite{4B,Tr,PRTr} may be presented in the form
\be\textstyle
X^0=\tau-\frac TD\s,\qquad
X^1+iX^2=u(\s)\cdot e^{i\om\tau}.
\label{soltr}\ee
Here the values $\s_i$, $D=\s_3-\s_0$,
$\dot X^2(\tau,\s_i)=V_i^2$, $\tau^*-\tau=T$
are the constants, the complex function
$u(\s)=A_i\cos\om\s+B_i\sin\om\s$,
$\s\in[\s_i,\s_{i+1}]$ is continuous in
$[\s_0,\s_3]$. Owing to Eqs.~(\ref{qq}) six complex constants $A_i,B_i$
are proportional to $A_1$ (here we put $\s_1=0$)
\begin{eqnarray*}
&A_0=A_1,\qquad B_0=\textstyle\frac12(h_1-iK)\,A_1,
\qquad B_1=-\frac12(h_1+iK)\,A_1,&\\
&A_2=\big[1+\frac12h_2d_1(2c_1-h_1d_1-iKd_1)\big]\,A_1,\quad
B_2=-\frac12\big[2h_2c_1^2+(1-h_2c_1d_1)(h_1-iK)\big]\,A_1,&
\end{eqnarray*}
where $|A_1|^2=(1-V_1^2)/\om^2=2T/(KD\om^2)$ and we use the following
notations:
\begin{eqnarray*}
&K=2d_2d_0^{-1}d_1^{-1}\sin\om T/(G_2G_3-1),\quad
d_i=\sin\om(\s_{i+1}-\s_i),\quad
c_i=\cos\om(\s_{i+1}-\s_i),&\\
&G_i=(h_id_{i-1}d_i-d_{i-1}c_i-c_{i-1}d_i)/d_{i+1},\qquad
h_i=\om m_i\gamma^{-1}|V_i|^{-1}.&
\end{eqnarray*}
The notations $c_i$, $d_i$, $h_i$, $G_i$ are cyclical: $c_{i+3}\equiv c_i$,
$d_{i+3}\equiv d_i\dots$

Expression (\ref{soltr}) is the solution of Eq.~(\ref{eq}) and satisfies the
orthonormality, closure and boundary (\ref{qq}) conditions if the
parameters are connected by the relations \cite{PRTr,ClTr}
\begin{eqnarray*}
&2\cos\om T=G_1+G_2+G_3-G_1G_2G_3,\quad
D/T+T/D=(K^2+4+h_1^2)/(2K),&\\
&(G_{i+1}-G_i)\,d_i=(G_iG_{i+1}-1)(d_{i-1}c_{i+1}-c_{i-1}d_{i+1}).&
\end{eqnarray*}
For given $m_i$, $\gamma$ and a parameter measuring the rotational speed
(for example $\om$) one can find all mentioned values by solving these
equations.

A string position for this state (a section $t={}$const of the
surface (\ref{soltr})) is the curve composed of three segments of a
hypocycloid joined in three points (the quark positions).
Hypocycloid is the curve drawing by a point of a circle that is rolling
inside another fixed circle with larger radius.

The energy and the angular momentum of this state are \cite{Tr,PRTr}
\be\!\!
M=\gamma D\Big(1-\frac{T^2}{D^2}\Big)+\sum_{i=1}^3\frac{m_i}{\sqrt{1-v_i^2}}+
\Delta M,\;\;
J=\frac1{2\om}\bigg[\gamma D\Big(1-\frac{T^2}{D^2}\Big)
+\sum_{i=1}^3\frac{m_iv_i^2}{\sqrt{1-v_i^2}}\bigg]+S.\!
\label{MJtr}\ee
where the quark velocities
$v_i=\sqrt{Td_{i-1}d_i(G_{i-1}G_{i+1}-1)\big/(Dd_{i+1}\sin\om T)}$.

A set of topologically different configurations of the system
is classified in Refs.~ \cite{PRTr,ClTr}. It is described, in particular, by integer
parameters $n$ and $k$ which (if there are no singularities
$\dot X^2=0$ in the segment $\s\in[\s_0,\s_1]$) are determined by the following way:
$$n=\lim\limits_{m_i\to0}\frac D{\s_1-\s_0},\qquad
k=\lim_{m_i\to0}\frac T{\s_1-\s_0}.$$
Here $k=n-2,\,n-4,\dots,-n+2$.
In the case $n=2$, $k=0$ (``quark-diquark" state) two quarks coincide,
one of $d_i$ equals 0, $T=0$. For this state the world surface (\ref{soltr})
reduces to (\ref{sol}) and the model ``triangle" --- to the q-qq one with the
tension $2\gamma$ (rectilinear segment is the particular case of the
hypocycloid). In the state $n=3$, $k=1$ (simple state) the string
has a (curvilinear) triangle form, for other $n$ and $k$
(exotic states) it has more complicated form with the cusps moving at the
speed of light \cite{PRTr,ClTr}.

The simple states of this model are used below for describing the RT.
It was shown in Ref.~\cite{stabTr} that they are stable (small disturbances
don't grow unlike in the q-q-q case \cite{lin}), in particular,
with respect to transformation into the ``q-qq" state with $n=2$, $k=0$.

In the two variants of describing RT with the help of string models
in Refs.~\cite{Ko} and \cite{4B} the spin-orbit correction
to the energy
\be
\Delta M_{SL}=\sum\limits_i\beta(v_i)(\mbox{\boldmath$\om s$}_i)
\label{corr}
\ee
were used in the two different forms. In Ref.~\cite{Ko} the expression
$\beta(v_i)=-\big[(1-v_i^2)^{-1/2}-1\big]$ is due to the Thomas
precession of the quark spins. It is obtained under the assumption that
in the quark rest frame the field is pure chromoelectric \cite{Ko,AllenOVW}.
The alternative assumption about pure chromoelectric field in the rotational
center rest frame results in $\beta(v_i)= 1-(1-v_i^2)^{1/2}$
\cite{4B,AllenOVW}.

The ultrarelativistic asymptotic behavior of the dependence $J(M)$
for the q-qq or mesonic (m), Y and $\Delta$ configurations has the form
\be
J\simeq\alpha'M^2-\nu M^{1/2}\sum_{i=1}^Nm_i^{3/2}
+\Delta J,\qquad v_i\to1,
\label{asymp}\ee
where $\displaystyle\alpha'=\frac1{2\pi\gamma}\cdot\left\{
\begin{array}{lr}1, &\mbox{m},\\
2/3, &\mbox{Y},\\  n/(n^2-k^2), &\Delta,\end{array}\right.\quad
\nu=\frac1{\sqrt\pi\,\gamma}\cdot\left\{
\begin{array}{lr}2/3, &\mbox{m},\\
(2/3)^{3/2}, &\mbox{Y},\\
\frac{\sqrt2\,n}3(n^2-k^2)^{-3/4}, &\Delta.
\end{array}\right.$
The term $\Delta J$ equals $\Delta J=\sum_{i=1}^Ns_i\big[1-\beta(v_i)\big]$
for the spin-orbit correction (\ref{corr}) \cite{4B} and
$\Delta J=a_n-1/2$ for the string model \cite{Solov}.

The Regge slope $\alpha'\simeq0.9$  GeV${}^{-2}$ is close
for mesons and baryons. So the effective value of string tension $\gamma$
is to be different for the baryon models Y, ``triangle" (simple states)
and q-qq. The following values of the parameters are used here and in
Ref.~\cite{4B} (for this values of $\gamma$ the string baryon configurations
q-qq, Y and $\Delta$ result in very close RT):

\begin{center}
Table 4. Input parameters in the string model \cite{4B}\\
\smallskip

\begin{tabular}{|l|l|l|} \hline
$\gamma=\gamma_{q-qq}=0.175$ GeV${}^2$ \rule[-3mm]{0mm}{8mm} & $\gamma_Y=\frac23\gamma$&
$\gamma_\delta=\frac38\gamma$\\ \hline
$m_u=m_d=130$ MeV &  $m_s=300$ MeV & $m_c=1600$ MeV \\ \hline
\end{tabular}
\end{center}

In Refs.~\cite{Ko} for describing baryon (and meson) RT with the help of the
quark-diquark model the values $m_{ud}=340$ MeV, $m_s=440$ MeV are used.
But the Thomas precession term in the correction (\ref{corr}) results in
the diquark mass strongly depending on the diquark spin state (this mass
varies from 220 to 550 MeV).
On the other hand in Ref.~\cite{Solov} the current quark masses
are used for describing the mesonic RT.

In this paper we use the string models (q-qq, Y and $\Delta$ for baryons) with
the parameters in Table 4 and the spin-orbit correction (\ref{corr}) with
$\beta(v_i)= 1-(1-v_i^2)^{1/2}$ for describing parent RT.

The results for light baryons (N and $\Delta$ RT) are shown in Figs.~4 and 5.
The particle data are taken from the PDG98 issue
\cite{PDG}. In Fig.~4a the N-baryon states with
$J^P=1/2^+$, $3/2^-$, $5/2^+$, $7/2^-\dots$ are placed and described
by our two models as two different RT. The red dotted line shows
the potential model \cite{Inop} predictions for the RT with positive parity
$p\;1/2^+$, $N(1680)\;5/2^+\dots$; blue dash-dotted line corresponds to the
blue marked states with negative parity. The same notations are used in
Figs.~5\,--\,8.

The dependence $J=J(M^2)$ for the string model ``triangle" and
``three-string" is shown in Figs.~4\,--\,7 as black solid lines
(under conditions in Table 4 these curves for the both configurations
practically coincide). Dots correspond to the quark-diquark configuration.
The results of the latter model are very close to that of two previous ones.

String hadron models \cite{Ko}\,--\,\cite{4B} at the modern stage are
(semi)classical and do not describe the isospin and parity. So in this paper
we use only the spin projection $S=\sum s_i$ (here $J=L+S$)
for modeling various RT. For example, in Fig.~4a the N states with
positive and negative parity may be described as two different
RT\footnote{One can describe all states in Fig.~4a as one RT \cite{Ko}
with the help of various string models. But such an approach requires
some (depending on parity and isospin) corrections to energy and
is not applicable to other trajectories in Figs.~4b, 5.}
in all (potential and string) models.
In the string models the spin states $S=1/2$ and $S=-1/2$
are used for these two RT --- the curves $J=J(M^2)$ are close to
rectilinear and fit the majority of states, except for the nucleon
($L=0$) and the states with $M>3000$ MeV which need confirmation
\cite{Hendry} (they are omitted from the summary table \cite{PDG}).
But the latter baryons are described in the potential model
\cite{Inop} rather well.

For the $\Delta$-resonances in Fig.~5 the similar picture take place.
The states on the $\Delta$(1232) trajectory with positive parity
are supposed to have $S=3/2$.

In Fig.~4b the N-baryon states with $J^P=1/2^-$, $3/2^+$, $5/2^-\dots$
are described by two different RT in the potential model \cite{Inop}
or by one RT with $S=1/2$ in the string models.
Remind that the string models are not applicable to the states
with small $J$ or $L$. Otherwise, the potential model
\cite{Inop} is the most adequate for small $L$.
It describes the strongly nonlinear RT generated by N(1535) $1/2^-$
(Fig.~4b) and $\Delta$(1910) $1/2^+$ (Fig.~5b). The string models
don't predict such nonlinear behavior (convexity) at small $J$.

In Fig.~6 the results of the string models $\Delta$ and Y
(q-qq is close to them) for the strange baryons $\Lambda$,
$\Sigma$ and $\Xi$ are represented. The potential model \cite{Inop}
predictions are calculated for $\Omega$ baryons.
The description of the string models is the most adequate for the
particles, whose spin state may be interpreted as $S=\pm1/2$.
For the RT with $S=3/2$, for example, $\Sigma$(1385), $\Omega$(1670),
$\Delta$(1232) in Fig.~5 the curves lie higher the lowest states
with $L=0$ and $L=2$. For larger $J=L+3/2$ the correspondence
is better (Fig.~5) but we have no reliable strange $\Sigma$, $\Xi$
and $\Omega$ mesons with $J\ge11/2$. There are some reasons
of such results in the string models \cite{4B}: (a) they are not
adequate for low $L$, (b) the spin-spin interaction must be
taken into account at low $L$, (c) perhaps, the form of the
spin-orbit correction (\ref{corr}) needs some improvement.

In the charmed sector (Fig.~7) we know only a few states with
$J=3/2$ and only one of them --- $\Lambda_{c1}^+$(2625) ---
may be interpreted as the orbital excitation of $\Lambda_c^+$.
In Fig.~7 these particles are described by the q-qq, Y and $\Delta$
string configurations with the spin state $S=1/2$ and the summary
spin of the light quarks $s_u+s_d=0$ (the upper curves in Fig.~7)
similar to the strange $\Lambda$ RT in Fig.~6.
For the state $\Sigma_c$ the trajectories with $S=-1/2$ and $s_u+s_d=-1$
is suggested. For other charmed baryons with $L=0$ the considered
string models are unlikely adequate.

For any string baryon model the corresponding
meson-like model with the same (or corresponding) string tension,
effective quark masses, types of spin or other corrections must
describe the RT in the meson sector.
The results of this ``meson test" for the string model
\cite{4B} are represented in Fig.~8 for light and strange mesons
in comparison with the results of the string meson model \cite{Solov}.
The latter model contains more fitting parameters (a set of $a_n$
in addition to the common ones $\gamma$ and $m_i$) so it describes
some mesons better. But the results of the model \cite{4B} are
satisfactory for the majority of light and strange meson states.

\bigskip
\noindent{\large\bf Conclusion}
\medskip

During the last two decades, a number of authors have shed some light on the problem of RT in the meson, baryon and glueball sectors. Our approach in this paper is to analyze both {\it pure experimental} RT and predict properties of the RT in two different models: string and potential. This way we hope to understand better the fundamental property of RT --- its linearity and when it is {\it broken}.  In particular, our analysis of all available data from PDG98 \cite{PDG} reveals that the following RT are essentially nonlinear (see Table 5).

\begin{center}
Table 5. Slopes for nonlinear baryon, meson RT\\
($\al'$, average $\langle\al'\rangle$, mean square deviation $\s$ in GeV${}^-2$)\\
\smallskip

\begin{tabular}{|l|l|c|c|} \hline
RT for baryons& slopes $\al'$ for neighbor pairs &
 $\langle\al'\rangle$&
$\s=\sqrt D$\\ \hline
N $3/2^-$ parent & 0.80 1.02 0.702 0.45 & 0.74 & 0.20 \\ \hline
$\Delta$ 1/2+ parent & 3.53 1.20 0.40& 1.71 & 1.33 \\ \hline
N $1/2^-$ parent & 4.44 0.89            & 2.67 & 1.78 \\ \hline
N $1/2^-$ radial & 2.73 0.61            & 1.67 & 1.06 \\ \hline
N $3/2^-$ radial & 1.72 0.70            & 1.21 & 0.51 \\ \hline
$\Delta$ $1/2^-$ parent& 1.82 0.98 1.11 0.60 0.34&0.97 & 0.51 \\ \hline
\hline
RT for mesons & \ \ \ slopes $\al'$  & $\langle\al'\rangle$ &
$\s $\\ \hline
f$(0^{++})$ parent & 3.00 0.78 0.94     & 1.58 & 1.01 \\ \hline
K$(0^-)$ parent & 0.73 1.14 0.34 1.25 & 0.87 & 0.36 \\ \hline
$\rho-a_2$ parent&0.87 0.90 0.82 0.69 2.08& 1.07& 0.51 \\ \hline
$\Upsilon$ radial&0.09 0.15 0.21 0.16 0.30& 0.18& 0.070 \\ \hline
$\chi_b$(1P) parent& 1.58 2.37 & 1.97 & 0.40 \\ \hline
$\chi_b$(2P) parent & 2.02 3.66 & 2.84 & 0.82 \\ \hline
J/$\Psi$ radial & 0.25 1.60 2.11 1.03 0.46 & 1.09& 0.69 \\ \hline
f${}_0$ radial &1.09 2.68 0.55 6.1 1.68 & 2.42 & 1.97 \\ \hline
f${}_2$ radial &{\small2.39 2.48 4.27 1.64 1.82 4.95 1.9 1.42 5.13}& 2.88 &1.39 \\ \hline
$\chi_c$(1P) parent & 1.54 3.15 & 2.34 & 0.80 \\ \hline
\end{tabular}
\end{center}

So, in total we have 6 baryon and 10 meson experimental RT, which we consider as {\it essentially nonlinear}.  We definitely witness the fundamental fact, that RT are not linear as a rule --- it depends on intrinsic quark-gluon dynamics.
One can observe nonlinearity for various RT in all current resonance region
--- for low or large momenta\footnote{But the data for the upper states with
the largest $J$ are not reliable in some cases \cite{PDG}.}.
Slopes for many experimental RT strongly differ from the standard $\al'\simeq0.9$ GeV${}^{-2}$ and vary significantly along the given RT ($\s\sim1$ GeV${}^{-2}$).

Now we return to theoretical predictions and interpretations of RT for baryons, mesons and glueballs.  In the series of papers \cite{Step}, the authors introduced a new procedure for the solution of the SE for mesons, which is based on the expansion in the Planck constant $\hbar$. In contrast to a number of papers, which employed a WKB approximation, the authors of \cite{Step} assumed that their procedure will work for large as well as for small values of the radial quantum number $N_r$, and that this method is complementary to the WKB method. The authors investigated $\rho-a_2$, J$/\Psi$ and $\Upsilon$ parent and daughter RT and show that all usually employed potentials lead to nonlinear RT in the resonance region. However, authors \cite{Step} did not consider baryonic systems and they did not show that the model adequately describes the {\it experimental} meson spectrum, which diminished the value of their papers \cite{Step}.

The authors of \cite{SimFR} investigated the baryon's RT in a relativistic approach, based on the method of vacuum correlators (MVC). While deriving the solutions of the dynamic equations, the authors neglected the spin, isospin, Pauli principle and one-gluon exchange potential, and thereby deduced that the RT were linear and found the N, $\Delta$ mass spectrum up to and including $K = 6$. They found the slope of baryonic RT to be equal to the slope of mesonic RT ($\al'=0.75$ GeV${}^{-2}$). It is interesting that authors compare the results of this approach with the previous paper of one of the authors (F.R.) \cite{FR}, which was based on the nonrelativistic SE with the power law confinement potential $r^{2/3}$.  The authors of \cite{SimFR} came to the conclusion that both methods gave very close spectra and linear RT. However, the above-cited authors \cite{Step} sharply criticized the paper \cite{FR} for misrepresenting the real picture in the resonance region, where the trajectories have a nonlinear character.  Simonov, the leader of method \cite{SimFR}, lends support to our thesis in his collaboration with new authors; they very recently have noticed a {\it weak nonlinearity} of the mesonic RT \cite{Sim} using the same approach (MVC). Their observation of the nonlinearity intersects that of Olsson et al
\cite{Olsson}, who analyzed the mesonic RT in the relativistic flux tube model.

J.Dey et al \cite{Dey} described RT for mesons and baryons in $q$-deformed model and
found a strong effect of nonlinearity for these RT, for rotational and radial excitations,
including daughters. They study $\pi-b$, $\rho-a$, $K$, $K^*$, $\varphi$, $\omega-f$, $\Upsilon$,
 $\Psi$ meson families and N, $\Delta$, $\Lambda$, $\Sigma$ baryon families,
computing spectra up to $J=17/2$.

In his seminal paper \cite{t'H} 't Hooft developed an $1/N_c$ relativistic,
toy model for mesons, and he got a RT which possess some nonlinearity:
``... deviations from the straight line are expected near the origin as a consequence
of the finiteness of the region of integration and the contribution of the mass terms..."
Mostly, the results of numerical analysis are described by the following WKB
mass formula with nonlinear logarithmic term
\be M^2 (n)\;\to\; \pi^2 n + (\al_1+\al_2 )\log n + C^{st} (\al_1,\al_2  )  .
\label{Msq}\ee

Iachello et al \cite{Iach} have noticed nonlinearity in the experimental
masses for $\Psi$, $\Upsilon$ families  ``...We next plot in Fig.~13 and 14,
the ``vibrational" RT for the $\Psi$ and $\Upsilon$ families.
We note that experimental masses no
longer fall exactly linear on trajectories, but their trajectories are {\it slightly bent},
a feature also anticipated in the 't Hooft calculation. The deviation of the trajectories
from linearity may indicate a breaking of the $SO(4)$ symmetry. Another possible
explanation is that the effective value of $N$, in our formula appropriate for heavy mesons,
is not $N\to\infty$ ($N=100$ used in Figs. 13 and 14), but much smaller. This may be
due to coupling with break-up channels that effectively terminate the rotational and
vibrational bands. The $U(4)\ni O(4)$ gives the following dependence on $N$:
\be M^2(\upsilon)  =  M_0^2 - 4(N+1)\,A\{\upsilon-\upsilon^2 /(N+1)\},
\label{Msq2}\ee
with $N\simeq20$, we find the ``vibrational" trajectories to be significantly bent,
so as to actually describe the vibrational spectra of $\Psi$ and $\Upsilon$  observed so far.
More experimental work is needed to extend these studies of the excited
vibrational states, to clarify the picture" (\cite{Iach}, p. 910).

We want to stress that few groups of authors simultaneously noted and
discussed the nonlinearity of the experimental RT for $J/\Psi$ and $\Upsilon$
families and its interplay with current theoretical quark models, namely Iachello
\cite{Iach}, Semay \cite{SemayC}, Stepanov \cite{Step}, Dey   \cite{Dey} and Sergeenko
 \cite{Sergeenko}.

In their series of papers Basdevant et al \cite{Basd} employ semirelativistic
hamiltonian and the HF method to describe spectra and RT of mesons and baryons.
When authors have tried to fit simultaneously mesonic and baryonic spectra, they found that
Regge slopes for mesons and baryons should be different. Authors describe
experimental spectra of $\Delta$, $\Sigma$, $\Xi$ and $\Omega$ up to
$J = 19/2$ with fair quality.

Johnson et al \cite{JohnsonN} used a semiclassical relativistic model for
the orbital spectra of mesons, based on the assumption of a universal,
flavor-independent linear confining interaction. They considered {\it yrast lines}
for the families $\pi$, $\eta$, $\eta'$, $\rho$, $\omega$,  $K$, $K^*$, $\varphi$,
$D$, $D^*$, $D_s$, $ D_s^*$, and, in particular, noted that $\pi$ and $K$ families
do not lie on linear Chew-Frautschi plots. Johnson also found that the $K^*$
family and $\rho$ family indeed have different slopes.

Durand et al \cite{Durand} also obtained varying slopes, describing
spin-averaged spectra of strange, charmed and bottom mesons in the
Bethe-Salpeter approach. Their range of variation for the slopes is
very similar to that of the $\hbar$-expansion technique \cite{Step},
which is  kind of a consistency check, because the two models started from
different dynamical origins.

Iachello \cite{Iach89} analyzed parity doubling phenomena in the meson and baryon sectors.
He has found out that parity doubling (PD) definitely takes place at low momenta,
maximally pronounced at $J = 5/2$, both experimentally and theoretically.
Then, only at higher momenta linear RT appeared.
``... One also observes that, at {\it high angular momenta},
parity doubling gradually disappears and linear
Regge trajectories with slope identical to that of
the meson spectra begin to appear..."
(\cite{Iach89}, p.38c, Figs. 10, 11). It is marvelous,
that our model \cite{Inop}
practically reproduces independently Figs. 10, 11
from \cite{Iach89}, while employing NRQM with HF method.

Sergeenko \cite{Sergeenko} derived a simple interpolating formula for the square of the quarkonium mass and an {\it analytic} expression for the RT in a whole region of both light and heavy quarkonia on the basis of the consideration of two asymptotics for the QCD inspired interquark potential. He found effects of nonlinearity for the $\Psi$ and $\Upsilon$ families RT and also the effect of decreasing slope with the increasing quark mass.

Very recently Filipponi et al \cite{Filip} built a simple phenomenological model which attempted to describe analytically the quark-mass dependence in the mesonic RT for all flavors. These authors definitely got the gross feature of decreasing slope with increasing quark mass for RT in $u$, $d$, $s$, $c$, $b$ - sector. The major drawback of the papers \cite{Filip} is that the authors did not account for $J$-dependence of the slopes. In spite of this oversimplification, the trend of their results roughly resembled that of our Table 3 for median values $\langle\al'\rangle_i$ over the whole $J$ interval.

Martin \cite{Martin} also considered parent RT for mesons and baryons in WKB-approach and has derived simple {\it analytical} formulas for flavor-dependent slopes, which are decreasing with increasing quark mass. Martin's formulas are very different from the Filipponi's results \cite{Filip}, and he also noticed nonlinearity effects for RT in the resonance region.

During the last few years group of T.Goldman obtained few new and very unusual results on RT \cite{Goldman,Burak}. Authors \cite{Goldman} used an unquenched lattice potential and compute the spectrum of the bottomonium system. They demonstrate numerically that the effect of color screening is to produce {\it termination} of {\it nonlinear} hadronic parent and daughter's RT. These results are supported by recent lattice calculations, which observe {\it string breaking} \cite{Karsch}.
It would be interesting to study the effect of color screening on light
quarkonia and baryon's RT.

Burakovsky \cite{Burak} constructed
generalized string model for RT with the string tension $\gamma$
depending on string point with broken reparametrizational symmetry.
 This model produces nonlinear Regge trajectories, which are in many
cases {\it analytical}, such as square root, logarithmic and hyperbolic trajectories.

L.\,D. Soloviev in Ref.~\cite{Solov96} suggested another string
model with nonlinear RT admitting {\it any growth rate} for the
trajectories. This power-law growth is limited by the
exponent 3/2. Linear growth is typical for some special cases of quark interaction. The nonlinearity of the RT in the model \cite{Solov96}
also results from some modification (dependence of the Lagrangian
on scalar curvature) of the standard string action (\ref{S}).
In standard string models \cite{BN} the RT are nonlinear
only for small $J$.

There are very few papers considering RT for glueballs. Goldman et al \cite{GoldPRD}  used the ``glueball dominance" picture of  the mixing between $q\bar q$ mesons of different hidden flavors and established new glueball-meson mass relations, which predict the following glueball masses:
$M(0^{++}) \simeq1.65\pm 0.05$ GeV,
$M(1^{--})\simeq 3.2 \pm 0.2$ GeV, $M(2^{-+})\simeq 2.95\pm 0.15$ GeV,
$M(3^{--})\simeq 2.8 \pm 0.15$ GeV. Their results are consistent with
(quasi)linear RT for glueballs with $\al'_G \simeq 0.3\pm 0.1$ GeV${}^{-2}$.

Yamada et al \cite{stringlue} investigated RT of gluonia (2g)
and hybrid mesons in the extended covariant oscillator quark model
(COQM), and estimated glueball slope $\al'_G \simeq 0.26$ GeV${}^{-2}$.
Authors predicted that the orbitally excited multiplets of gluonia with
$L = (1,2,3)$ lie in the mass region of (2.5 - 3.0 GeV, 3.2 - 3.6 GeV,
3.7 - 4.1 GeV). Their results are in fair agreement with lattice predictions.

The relativistic quantum string model by L.\,D. Soloviev \cite{Solov} describing
reasonably well meson spectrum also predicts glueball RT with
$J^{PC}=0^{++}$ --- f${}_0$(1370) and f${}_0$ (1500), which must lie
on the following trajectories
$$\sqrt{J(J+1)}  =  m_2 /(4\pi\gamma) + a_g.$$

The results of our string and potential model fits and predictions for
baryon and meson spectra and RT reveals distinctive feature ---
RT in many cases are {\it nonlinear} functions of $J$.  This {\it fundamental}
feature is in accord with analysis of {\it pure experimental} RT from PDG98 (Table 5), and with predictions of dozens different quark models,
reviewed in this paper.  Regge trajectories for mesons and baryons
{\it are not} straight and parallel lines in general in the {\it current
resonance  region} both experimentally and theoretically, but
{\it very often} have appreciable curvature, which is flavor-dependent.

\end{document}